\documentclass{optica-article}

\journal{opticajournal} 

\articletype{Research Article}

\usepackage[separate-uncertainty = true, multi-part-units = single]{siunitx}
\DeclareSIUnit{\ppm}{ppm}
\DeclareSIUnit{\ppb}{ppb}
\DeclareSIUnit{\ppt}{ppt}
\DeclareSIUnit{\sccm}{sccm}
\DeclareSIUnit{\pixel}{px}
\DeclareSIUnit{\bar}{bar}
\DeclareSIUnit{\rydberg}{Ry}
\DeclareSIUnit[quantity-product = ]\percent{\char`\%}
\DeclareSIUnit{\inch}{in.}
\DeclareSIUnit{\grooves}{G}

\usepackage{fp}

\newcommand{\SiO}{SiO\textsubscript{2}}

\newcommand{\AlO}{Al\textsubscript{2}O\textsubscript{3}}

\newcommand{\NtwoO}{N\textsubscript{2}O}

\newcommand{\COtwo}{CO\textsubscript{2}}
\newcommand{\CO}{CO}
\newcommand{\TaO}{Ta\textsubscript{2}O\textsubscript{5}}

\newcommand{\aSi}{a-Si}

\usepackage{lineno}

\begin{document}

\title{Ion-Beam-Sputtered Mid-Infrared Coatings for Hybrid Supermirrors}

\author{Lukas W. Perner,\authormark{1,*} Valentin J. Wittwer,\authormark{1,2}, Gar-Wing Truong,\authormark{3,4} Seth B. Cataño-Lopez,\authormark{3} Garrett D. Cole,\authormark{3,5} and Thomas Südmeyer\authormark{1}}

\address{\authormark{1}Laboratoire Temps-Fréquence, Institut de Physique, Université de Neuchâtel, Avenue de Bellevaux 51, CH 2000 Neuchâtel, Switzerland\\
\authormark{2}UTOM AG, Ostringstrasse 10, CH 4702 Oensingen, Switzerland\\
\authormark{3}Thorlabs Crystalline Solutions, 114 E Haley St., Suite G, Santa Barbara, CA 93101, USA\\
\authormark{4}Current address: Freedom Photonics, 41 Aero Camino, Santa Barbara, CA 93117, USA\\
\authormark{5}Current address: Wyant College of Optical Sciences, University of Arizona, 1630 E University Blvd., Tucson, AZ 85721, USA
}

\email{\authormark{*}lukas.perner@unine.ch}

\begin{abstract*} 
We report on the development of low-loss ion-beam-sputtered (IBS) mid-infrared coatings for hybrid supermirrors. Two highly reflective designs were realized: HR1, a 4-period \aSi{}/\SiO{} DBR with an \AlO{} bonding layer, and HR2, a 6-period \aSi{}/\TaO{} DBR with an \aSi{} terminating layer for bonding. Combined with a GaAs/AlGaAs crystalline mirror, HR2-based hybrids yielded a total loss of \qty{9.3}{\ppm} and excess loss of \qty{6.8}{\ppm} at \qty{4.45}{\um}, with cavity finesse up to \num{396000}. For the first time, we used IBS-deposited \aSi{} directly as a bonding layer, verified its sub-angstrom roughness, and demonstrated excellent optical performance. This establishes a clear path toward scalable coatings for longer mid-infrared wavelengths, building on prior results in Nat. Commun. {\bfseries 14}, 7846 (2023).
\end{abstract*}

\section{Introduction}
\label{sec:Introduction}
Advanced low-loss optical coatings for the mid-infrared (MIR) spectral region are an indispensable tool for a wide variety of applications, ranging from various types of MIR lasers, such as QCLs~\cite{bai_optimizing_2011, wysocki_widely_2008} or OPOs~\cite{pecile_record-high_2024, adler_phase-stabilized_2009}, to ubiquitous passive devices such as antireflection-coated (AR-coated) lenses, partially reflecting or dichroic beamsplitters, and various types of highly reflective (HR) mirrors~\cite{fleisher_optical_2017, terabayashi_optical_2017, mccartt_room-temperature_2022, kaariainen_optical_2019, truong_mid-infrared_2023-1, winkler_mid-infrared_2021-1}.
In particular, HR multilayers are a key component for broad- and narrow-band cavity-enhanced spectroscopy techniques~\cite{romanini_introduction_2014,zhao_doppler-free_2020,delli_santi_biogenic_2021,truong_mid-infrared_2023-1, giusfredi_saturated-absorption_2010, giusfredi_theory_2015, zhao_frequency_2021}, where sensitivity immediately benefits from reduced excess optical loss, that is scatter and absorption $S+A$, of cavity mirrors. Some of the most promising spectroscopic targets are at wavelengths of \qty{4}{\um} and beyond, where experiments can probe the fundamental transition of molecules such as \CO{}, \COtwo{}, and \NtwoO{}. The absorption linestrength of these fundamentals are much stronger than their overtones in the near-infrared (NIR) and visible (VIS) region~\cite{shumakova_short_2024}.

However, the performance of HR coatings in the MIR has been lagging behind their counterparts in the VIS and NIR, where finesse values over 1 million, corresponding to a per-mirror total loss $T+A+S$ of approximately \qty{3}{\ppm}, have been achieved~\cite{jin_micro-fabricated_2022}. In particular, multilayers deposited by various physical vapor deposition (PVD) processes --such as evaporation (thermal, e-beam; optionally ion-assisted), as well as magnetron or ion-beam sputtering (IBS)-- traditionally suffer from high excess optical loss in the MIR. This is inherent to the deposition process, the materials used, or both.
Furthermore, in the past, process optimization in coating technology was often focused on oxides, which are difficult to use at longer wavelengths due to their increased absorption from infrared vibrational modes, with prominent bands around \qty{9}{\um} for \SiO{} and \qty{11}{\um} for \TaO{}~\cite{franta_optical_2016,franta_wide_2025}, making the latter more suitable for longer-wavelength applications.

Nonetheless, considerable efforts have been made to improve all-amorphous coatings for the MIR region, with commercial HR coatings achieving total loss $1-R$ below \qty{30}{\ppm} (parts per million, \num{e-6}), e.g., with a finesse of \num{114000} at a wavelength $\lambda = \qty{4.5}{\um}$ in Ref.~\cite{kaariainen_optical_2019}. Furthermore, studies such as Refs.~\cite{chen_comparison_2020, habel_group_2016} generate valuable insights concerning the suitability of various materials and deposition methods.

In addition to all-amorphous multilayers, monocrystalline substrate-transferred GaAs/AlGaAs heterostructures have recently emerged as an excellent alternative for HR coatings in the MIR region~\cite{cole_high-performance_2016,winkler_mid-infrared_2021-1,truong_mid-infrared_2023-1}.
Such all-crystalline distributed Bragg reflectors (DBRs) are grown via molecular beam epitaxy (MBE) and yield extremely low $S+A$ in the MIR. However, their monocrystalline nature requires materials with a closely matching lattice constant. While several such lattice-matched material combinations have been explored for DBRs directly on the MBE growth wafer in the past~\cite{heiss_epitaxial_2001}, currently only GaAs/AlGaAs-based substrate-transferred MIR supermirrors have been demonstrated~\cite{cole_high-performance_2016,cole_laser-induced_2018,winkler_mid-infrared_2021-1,truong_mid-infrared_2023-1}, as they can be grown with sufficiently low background doping and growth defects. This limits free-carrier absorption at longer wavelengths and makes them suitable for ultra-low loss optical devices with \unit{\ppm}-level excess loss~\cite{cole_high-performance_2016}. However, these beneficial properties come at the cost of the relatively low refractive index contrast of GaAs and AlGaAs, which leads to the need of a high number of quarter-wave layer pairs to obtain DBRs with optimal reflectivity $R$ for cavity-enhanced methods~\cite{perner_simultaneous_2023,truong_mid-infrared_2023-1}. Especially in the MIR, this leads to relatively thick coatings and narrow mirror stopbands. Finally, the maximum thickness of MBE-grown monocrystalline structures at an acceptable defect density is limited. This, in turn, impedes scaling of all-crystalline GaAs/AlGaAs supermirrors to center wavelengths (CWL) $\lambda_0$ well above \qty{5}{\um} when aiming for single-\unit{\ppm}-level excess loss.

In fact, the most recent iteration of all-crystalline mirrors with a CWL around \qty{4.45}{\um} already employs a sandwiched design, where two MBE-grown half-mirrors are bonded to form a 44.5-period GaAs/AlGaAs DBR based on a bottom and top partial mirror of 22.25 periods each~\cite{truong_mid-infrared_2023-1}, with a previous design of 34.5 periods using the same approach~\cite{winkler_mid-infrared_2021-1}. The all-crystalline HR coatings presented in Ref.~\cite{truong_mid-infrared_2023-1} have a total thickness of about \qty{32.2}{\um} and a total loss $1-R_0=\qty{13.60(0.49)}{\ppm}$ (i.e., $R_0=\qty{99.99864(49)}{\percent}$), composed of $T_0=\qty{9.33(0.17)}{\ppm}$ transmission and $S_0+A_0=\qty{4.27(0.52)}{\ppm}$ excess loss, corresponding to a maximum finesse of \num{231000} at a center wavelength of $\lambda_0=\qty{4.45}{\um}$~\cite{truong_mid-infrared_2023-1}.

Recently, a new hybrid amorphous-crystalline coating paradigm was invented~\cite{cole_substrate-transferred_2025}. In this approach, the bottom crystalline half-mirror is  replaced by an amorphous multilayer structure. This subcoating is deposited on the optical substrate via ion-beam sputtering (IBS) prior to transfer of the crystalline DBR. In the design presented in Ref.~\cite{truong_mid-infrared_2023-1}, these hybrid mirrors allowed for a two-mirror cavity finesse of \num{409000}, corresponding to a per-mirror total loss $1-R_0=\qty{7.70(0.27)}{\ppm}$ (i.e., $R_0=\qty{99.999230(27)}{\percent}$) at the expense of slightly higher excess loss and lower transmission of $S_0+A_0=\qty{5.17}{\ppm}$ and $T_0=\qty{2.53}{\ppm}$, respectively. The main technological advantage is that the lower \num{22.25} periods of GaAs/AlGaAs in the all-crystalline design is replaced with only four periods of \aSi{}/\SiO{} (with an \AlO{} 1/8-wave cap to aid in bonding), reducing the overall coating thickness to approximately \qty{21.0}{\um} while also avoiding the double bonding of the all-crystalline approach (thereby increasing the yield of a single cost-intensive MBE growth run).

Compared to all-amorphous HRs, hybrid designs can tolerate slightly higher $S+A$ in the IBS-deposited layers, as the electric field at the CWL primarily samples the excess loss from the surface crystalline layers. Nevertheless, limiting the total loss to $1-R<\qty{10}{\ppm}$ relies on advancements in both the crystalline and amorphous parts of the hybrid multilayer. Furthermore, the surface quality of the IBS films must be excellent, maintaining sub-\unit{\nm} RMS microroughness and negligible defect density to facilitate bonding of the crystalline mirror on top.

In this paper, we expand on the results presented in our recent conference contribution~\cite{perner_ion-beam-sputtered_2025} and present details of the amorphous HR and AR coatings developed for these hybrid mirrors. We present and characterize an alternative hybrid design that avoids the use of an \AlO{} adhesion layer, using only \TaO{} and \aSi{} layers. For that purpose, we describe the amorphous coating process, investigate the surface roughness of the terminating \aSi{} layer after IBS deposition, and compare it to a bare superpolished substrate as used for direct transfer bonding of MBE-grown DBRs. Finally, we explore the optical performance of all-amorphous AR and HR coating designs. Specifically, we show transmittance characterization of two HR coatings, HR1 and HR2, and an AR coating. Using hybrid mirrors based on HR2, we perform spectrally and spatially resolved cavity ringdown (CRD) measurements, exploring the performance of this new design.

\section{Coating Design}
\label{sec:Design}
As the first prototype hybrid mirrors were intended for proof-of-principle cavity-ringdown characterization and spectroscopy (see Ref.~\cite{truong_mid-infrared_2023-1} for details), they required both HR and AR coatings for optimal performance and efficient coupling of laser light in a linear two-mirror cavity. For the design of these coatings, we use refractive index $n$ data from Ref.~\cite{perner_simultaneous_2023} for GaAs/AlGaAs, Ref.~\cite{tatian_fitting_1984} for the Si substrate, and Ref.~\cite{ciddor_refractive_1996} for air, while the data for the \aSi{}, \SiO{}, \TaO{}, and \AlO{} thin films are generated in-house from test structures fabricated with the same IBS machine as the AR and HR structures.

\begin{figure}[htbp]
  \centering
  \includegraphics[width=13.335cm]{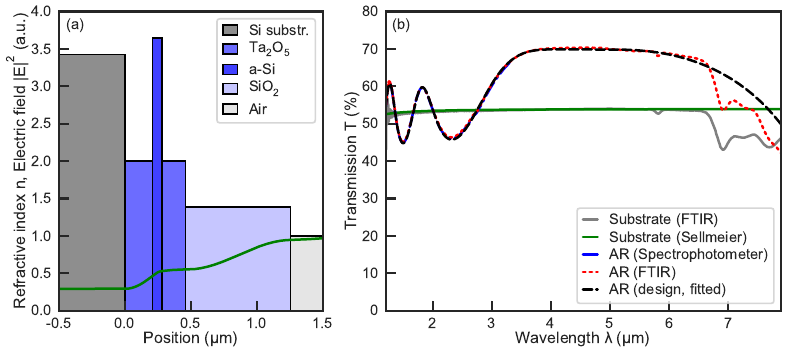}
\caption{
Details of the AR coating used on the backside of the hybrid mirrors: (a) AR design and simulated decay of the electric field intensity at \qty{4.45}{\um} (green line). (b) Measurements on the bare and coated witness pieces as well as best-fit curves based on the layer structure in (a). Deviations between model and measurement data at longer wavelengths are due to absorption in the substrate material of the witness pieces (see main text for details).
}
\label{fig:AR}
\end{figure}

The AR coating is designed for a reflectivity of $R_\text{AR} < \qty{0.1}{\percent}$ at near-normal incidence over a wavelength range of \qtyrange{4}{5}{\um}, around the mirror's CWL of $\lambda_0 = \qty{4.45}{\um}$. As shown in Fig.~\ref{fig:AR}(a), it is a three-material four-layer design employing \aSi{}, \TaO{}, and \SiO{}. In the final supermirrors, the AR coating is used to minimize Fresnel reflection losses on the back of the superpolished silicon substrates. Starting from a basic two-layer design (see, e.g., Ref.~\cite{macleod_thin-film_2018}) with an additional \aSi{} buffer layer, this structure is numerically optimized based on transfer matrix method (TMM) modeling in the above wavelength region using Python-based software developed in-house.

\begin{figure}[htbp]
  \centering
  \includegraphics[width=13.335cm]{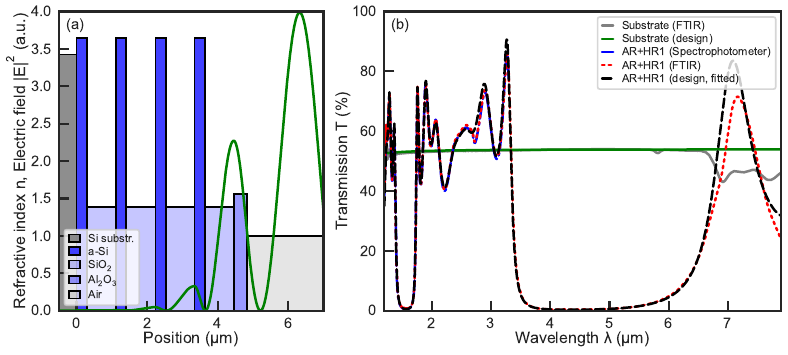}
\caption{
Details of the HR1 coating used for the hybrid mirrors in Ref.~\cite{truong_mid-infrared_2023-1}: (a) HR1 design and simulated decay of the electric field intensity at \qty{4.45}{\um} (green line). (b) Measurements on the bare and HR1 coated witness pieces (with backside AR coating) as well as best-fit curves based on the layer structure in (a). Deviations between model and measurement data at longer wavelengths are due to absorption in the substrate material of the witness pieces (see main text for details).
}
\label{fig:ARHR1}
\end{figure}

For the HR subcoatings, we developed two different structures, both centered around a classic quarter-wave DBR design: HR1, a 4-period \aSi{}/\SiO{} DBR with a terminating 1/8-wave \AlO{} cap, shown in Fig.~\ref{fig:ARHR1}(a). This design was extensively tested as part of the hybrid DBR presented in Ref.~\cite{truong_mid-infrared_2023-1}.

\begin{figure}[htbp]
  \centering
  \includegraphics[width=13.335cm]{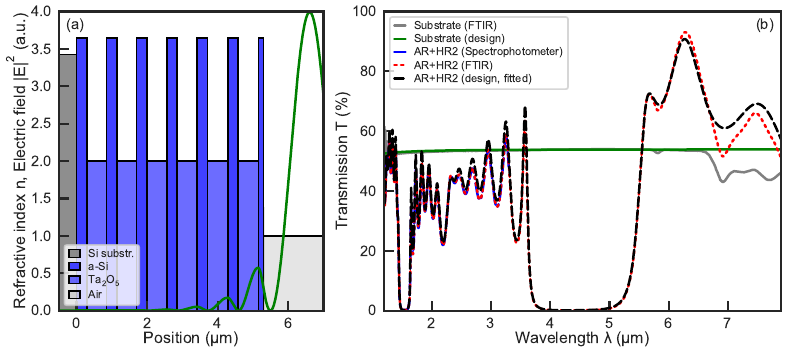}
\caption{
Details of the HR2 coating used for the hybrid mirrors characterized in this study: (a) HR2 design and simulated decay of the electric field intensity at \qty{4.45}{\um} (green line). (b) Measurements for the bare and HR2 coated witness pieces (with backside AR coating) as well as best-fit curves based on the layer structure in (a). Deviations between model and measurement data at longer wavelengths are due to absorption in the substrate material of the witness pieces (see main text for details).
}
\label{fig:ARHR2}
\end{figure}

In the present study, we characterize an alternative high reflector (HR2), a 6-period \aSi{}/\TaO{} DBR with a terminating 1/8-wave \aSi{} cap, shown in Fig.~\ref{fig:ARHR2}(a). It should be noted that this design uses only two materials in a strict alternation of high- and low-index layers.

\begin{figure}[htbp]
  \centering
  \includegraphics[width=13.335cm]{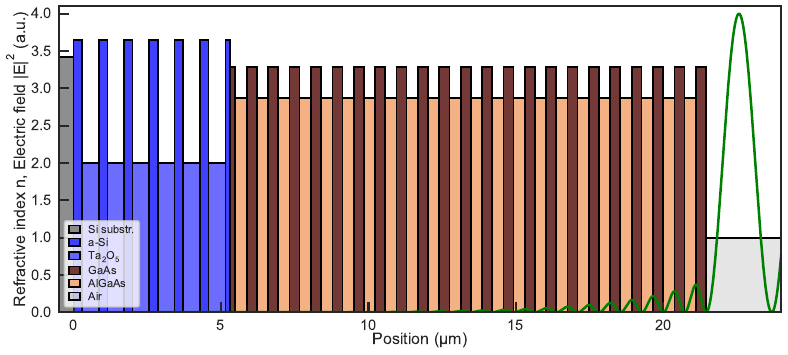}
\caption{
Multilayer design and simulated decay of the electric field intensity (green line) for the complete hybrid mirror based on HR2, as subjected to optical characterization in this study. The results of spectrally resolved measurements and spatial mapping of this design can be found in Figs.~\ref{fig:CRD} and~\ref{fig:HR2map}, respectively.
}
\label{fig:ARHR2cryst}
\end{figure}

\begin{figure}[htbp]
  \centering
  \includegraphics[width=13.335cm]{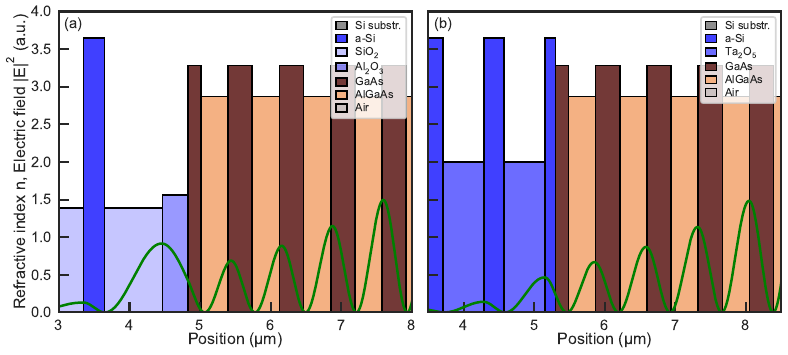}
\caption{
Comparison of the multilayer designs and simulated electric-field intensity profiles (green lines) for the complete hybrid mirrors based on (a) HR1 and (b) HR2. In both simulations, the incident field amplitude was multiplied by a factor of 500 compared to Fig.~\ref{fig:ARHR2cryst} and exhibits identical behavior within the crystalline region. In the HR1-based design, the field amplitude is significantly higher near the bonding interface and within the \AlO{} layer, which is unfavorable due to the resulting increase in optical loss. The HR2 design suppresses this enhancement by employing an \aSi{} bonding layer, leading to a markedly lower field around the bonding interface and within the first amorphous layers.
}
\label{fig:ARHR1cryst_ARHR2cryst_comp}
\end{figure}

In both HR1 and HR2, the terminating 1/8-wave caps match a similar GaAs 1/8-wave layer of the crystalline top coating. For HR1, we choose \AlO{} due to previous success in transfer bonding MBE-grown DBRs to \AlO{} surfaces. In contrast, HR2 ends with an \aSi{} layer. This benefits the complete hybrid mirror (shown in Fig.~\ref{fig:ARHR2cryst}), as the refractive indices of the respective \aSi{} and GaAs 1/8-wave caps of the amorphous and crystalline halves closely match each other. This, in turn, leads to a more favorable decay of $|E^2|$ in the vicinity of the bonding interface compared to HR1 (see Fig.~\ref{fig:ARHR1cryst_ARHR2cryst_comp} for details).

\section{Fabrication}
\label{sec:Fabrication}
For the complete hybrid mirrors, we deposited the respective amorphous designs from Sec.~\ref{sec:Design} on the curved (HR1, HR2, \qty{1}{\meter} radius of curvature) and flat (AR) sides of superpolished Si substrates (\qty{6.35}{\mm} thickness, \qty{25.4}{\mm} outer diameter, backside wedged at \ang{;30;}) via IBS deposition. Subsequently, the MBE-grown GaAs/AlGaAs multilayer was directly bonded on the respective HR sides. The details of this process and further characterization of HR1, as well as proof-of-principle spectroscopy can be found in Ref.~\cite{truong_mid-infrared_2023-1}. In each coating run, we also placed flat \qty{4}{\mm}-thick Czochralski-grown Si substrates (Crystran Inc.) as witness pieces, subsequently used for in-depth post-deposition characterization of the amorphous subcoatings, as presented in Sec.~\ref{sec:Characterization}.

All amorphous structures were deposited using a Navigator 1100 IBS coating machine (Cutting Edge Coatings GmbH) using xenon as a sputtering gas.
The \SiO{}, \TaO{}, and \AlO{} layers were reactively sputtered from high-purity metal targets (Si: p-doped, \qty{>99.9999}{\percent} purity; Ta: \qty{99.999}{\percent} purity; Al: \qty{>99.999}{\percent} purity) using \qty{70}{\sccm}, \qty{75}{\sccm}, and \qty{90}{\sccm} of oxygen, respectively.
In contrast, \aSi{} was sputtered without added oxygen from an undoped target (\qty{99.9999}{\percent} purity).
Before each run, the process chamber was evacuated to about \qty{1d-7}{\milli\bar} and kept below \qty{2d-3}{\milli\bar} throughout deposition.
Each coating run was completed without breaking vacuum and controlled via in-situ broadband optical monitoring (BBOM, wavelength range of \qtyrange{0.25}{2.2}{\um}) on IR fused silica monitoring glasses.
The substrates were preconditioned using the assist ion source of the IBS device (\qty{\sim 1}{\min} exposure to Ar ions to ablate \qty{\sim 2}{\nm}) prior to deposition to remove potential surface contaminants and kept at \qty{150}{\degreeCelsius} during deposition. Before fabricating the complete hybrid mirror, the IBS-coated optics were annealed for \qty{24}{\hour} at \qty{300}{\degreeCelsius} in a stainless steel hot air oven (Memmert GmbH) in ambient atmosphere to reduce absorption as compared to the as-deposited structure.

\section{Characterization}
\label{sec:Characterization}
To verify the performance of the IBS-fabricated samples, we subject flat Si witness pieces from the AR, HR1, and HR2 coating runs to transmission measurements by spectrophotometry (Agilent Cary 5000) and FTIR transmission spectroscopy (ThermoFisher Nicolet iS50). As the spectrophotometer is inherently calibrated, we rescaled the vertical axis of the FTIR measurements to coincide with the Cary measurements in the overlapping spectral range. This mitigates the effects of the highly refractive substrate material, typically leading to miscalibration of the vertical axis in FTIR transmittance measurements~\cite{hirschfeld_focal_1978}. For all subsequent curve-fitting exercises, we use data from the spectrophotometer and FTIR. To obtain accurate results, we first characterized a bare substrate from the same batch. As can be seen from the $T$ curves given in, e.g, Fig.~\ref{fig:AR}, the witness pieces exhibit excess absorption around \qty{5.8}{\um}, typical of oxygen absorption in Czochralski-grown silicon.

We subsequently characterize a witness piece with a single-sided AR coating, see Fig.~\ref{fig:AR}(b). This measurement was used to obtain the as-deposited layer thicknesses from best-fit results for a TMM model (based on the initial design). The best-fit model is in excellent agreement with the measured curves in the relevant spectral range. These updated values for the AR coating are also used in the characterization of witness pieces of the HR1 and HR2 coating runs, as their backside is coated in the same AR coating run.

For both HR coatings, we follow a similar approach, using witness pieces. Again, Figs.~\ref{fig:ARHR1}(b) and \ref{fig:ARHR2}(b) show excellent agreement between a best-fit model (varying the layer thicknesses) and measurements, owing to the BBOM during deposition.

\begin{figure}[htbp]
  \centering
  \includegraphics[width=13.335cm]{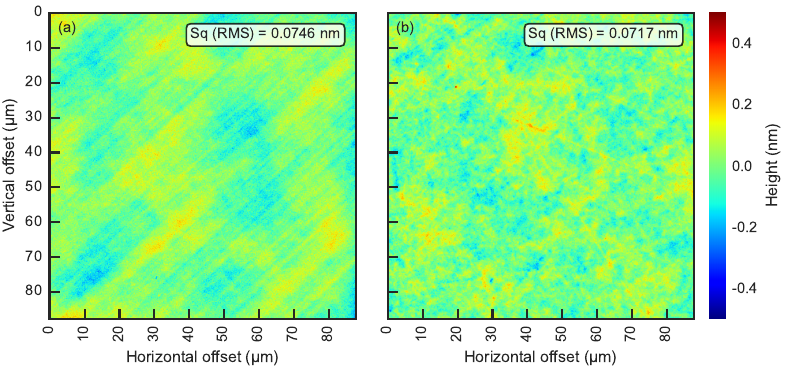}
\caption{
Surface roughness measurements of (a) a bare superpolished Si substrate and (b) the final \aSi{} layer of an annealed IBS-coated substrate. The measurements illustrate that the sub-angstrom roughness of the bare substrate is maintained through IBS depostion.
}
\label{fig:surf_roughness}
\end{figure}

For the present study, we fabricate a pair of hybrid mirrors using HR2. To verify suitability of the IBS-coated superpolished substrates for transfer bonding the MBE-grown GaAs/AlGaAs, we measure the surface roughness of the terminating \aSi{} layer after annealing. For that purpose, we subject a bare and an IBS-coated sample to coherence scanning interferometry in an optical surface profiler (Zygo NewView 9000). The results, shown in Fig.~\ref{fig:surf_roughness}, demonstrate that the initial roughness of the substrate is maintained throughout the IBS coating process, yielding sub-angstrom roughness for the bare and coated sample. In these measurements, we corrected for the \qty{1}{\m} ROC of the substrate by fitting a curved surface. Subsequently, we successfully fabricated a pair of HR2-based hybrid mirrors.

\begin{figure}
  \centering
  \includegraphics[width=13.335cm]{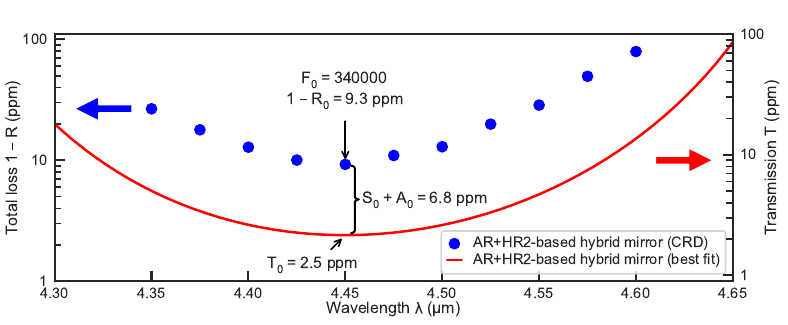}
\caption{
Spectrally resolved measurements of optical loss of the HR2-based hybrid mirror. Blue circles: total loss $1-R$ from CRD measurements. Solid red: as-grown transmission $T$ from a best-fit TMM model to broadband transmittance measurements. Annotations indicate the respective values at the mirror's CWL of $\lambda_0=\qty{4.45}{\um}$: inferred finesse $F_0$, measured total loss $1-R_0$, transmittance from TMM $T_0$, excess loss $S_0+A_0$ (scatter + absorption).
}
\label{fig:CRD}
\end{figure}

\begin{figure}
  \centering
  \includegraphics[width=7cm]{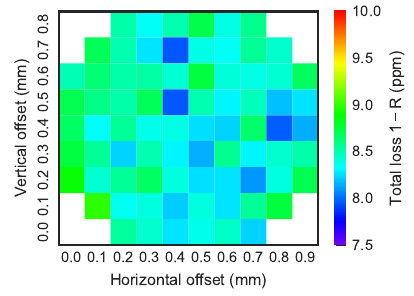}
\caption{
Spatially resolved total loss $1-R$ mapping of the HR2-based hybrid mirror. The value at each point represents the median value of 10 consecutive CRD measurements taken over an \qtyproduct{0.9 x 1.0}{\mm} area (sample pitch \qty{0.1}{\mm}). This measurement shows the excellent homogeneity of hybrid mirrors with an \aSi{} termination, maintaining the benefits shown for an \AlO{} cap in Ref.~\cite{truong_mid-infrared_2023-1}.
}
\label{fig:HR2map}
\end{figure}

We subject these prototype mirrors to extensive characterization of total loss $1-R$ and transmittance $T$ based on CRD and FTIR measurements, respectively. Details on the CRD apparatus can be found elsewhere~\cite{truong_mid-infrared_2023-1, winkler_mid-infrared_2021-1, truong_near-infrared_2019}. As shown in Fig.~\ref{fig:CRD}, spectrally resolved analysis at a single location yielded a two-mirror cavity finesse of \num{340000}, from which we infer a per-mirror total loss $1-R_0=\qty{9.3}{\ppm}$ and an excess loss of $S_0+A_0=\qty{6.8}{\ppm}$ at $\lambda_0 = \qty{4.45}{\um}$. The latter value is obtained by subtracting $T_0$ from the total loss. This transmittance at the CWL (well below the noise floor of direct FTIR measurements) is inferred from a TMM with layer thicknesses obtained from fitting the initial design to broadband FTIR data, similar to the approach shown in Figs.~\ref{fig:AR},~\ref{fig:ARHR1}, and~\ref{fig:ARHR2}. It is described in detail in Refs.~\cite{perner_simultaneous_2023,truong_mid-infrared_2023-1}. Scanning the beam center over an area of \qtyproduct{0.9 x 1.0}{\mm} in the CRD apparatus at $\lambda_0 = \qty{4.45}{\um}$, as shown in Fig.~\ref{fig:HR2map}, illustrates the excellent surface homogeneity, with an average $1-R=\qty{8.42}{\ppm}$ and a standard deviation of \qty{0.20}{\ppm} among the \num{78} probed spots. Some of the point-to-point variation in Fig.~\ref{fig:HR2map} is introduced by imperfect tip-tilt compensation of the automated mapping apparatus. We assume that the remaining nonuniformity can be attributed to the local variations in defect density, roughness, free-carrier concentration, and any bonding-related imperfections.

We maintain a total loss of \qty{<10}{\ppm} over the sampled area. One position exhibited $1-R=\qty{7.93}{\ppm}$, corresponding to a finesse as high as \num{396000}, exceeding the results of the spectrally resolved measurements in Fig.~\ref{fig:CRD}. Note that the individual squares represent the position of the beam center at the time of measurement. As the calculated $\tfrac{1}{e^2}$ beam radius on the mirror is \qty{0.629}{\mm}, the probed area extends substantially beyond the limits of this plot, representing about \qty{3}{\mm\squared} of the coating center.

\section{Conclusion and Outlook}
\label{sec:Conclusion}
In this study, we present results for the design, fabrication, and characterization of IBS-deposited AR and HR coatings, targeting operation in the mid-IR spectral range.

We realized two different HR designs for the purpose of this study, HR1, a 4-period \aSi{}/\SiO{} DBR with a terminating \AlO{} bonding layer and HR2, a 6-period \aSi{}/\TaO{} DBR with an \aSi{} bonding layer.
These HR coatings were used as part of novel amorphous-crystalline hybrid supermirrors, with hybrid mirrors based on HR1 shown to have record-high reflectivity around \qty{4.45}{\um} in Ref.~\cite{truong_mid-infrared_2023-1}. In the present study,
we performed spectrally resolved cavity-ringdown measurements on hybrid mirrors using HR2, yielding results comparable to those published for hybrid mirrors using HR1 in Ref.~\cite{truong_mid-infrared_2023-1}, with a two-mirror cavity finesse above \num{340000} at a center wavelength of $\lambda_0=\qty{4.45}{\um}$. Further 2D mapping of total loss in CRD revealed excellent homogeneity, with sub-\qty{9}{\ppm} total loss $1-R$ being maintained over an area \qty{>2.5}{\mm\squared} with minimal variations. In this mapping, few exceptional spots with finesse values as high as \num{396000} were found.
Together, these results suggest extremely low absorption in the IBS-manufactured part of both presented HR coatings. These hybrid mirrors exhibit excellent robustness under normal laboratory handling and can be treated like conventional high-reflectivity optics, making them a practical drop-in replacement for existing cavity-enhanced spectroscopy and metrology systems.

Furthermore, we realized a three-material four-layer AR coating for the back of the hybrid mirrors, facilitating their use in cavity-enhanced methods by reducing the coupling loss considerably.
We also performed surface roughness measurements on the capping \aSi{} layer of HR2, further illustrating the quality of the IBS coatings for hybrid mirror fabrication. This, together with the successful fabrication and characterization of the HR2-based hybrid mirror, demonstrates the suitability of \aSi{} as a bonding layer, making \AlO{} adhesion layers, as used in Ref.~\cite{truong_mid-infrared_2023-1}, unnecessary.

Therefore, we conclude that a combination of both HR designs, i.e., an \aSi{}/\SiO{} DBR with an \aSi{} 1/8-wave bonding layer, is a potential improvement for future designs, given the closely matched refractive indices of \aSi{} and GaAs. The \aSi{}/\SiO{} material pair benefits from an exceptionally high refractive index contrast, which reduces the required number of layer pairs for a given target reflectivity while simultaneously providing a wider stopband. Importantly, the 1/8-wave bonding cap is a remnant of all-crystalline designs, where two half stacks were bonded to form a complete mirror. In this study, we have demonstrated that direct bonding to \aSi{} is feasible, and thereby confirm the viability of the hybrid mirror approach using \aSi{} as a terminating layer. This opens the path for hybrid-specific MBE growth runs that omit the 1/8-wave cap altogether, enabling optimized hybrid coatings: \aSi{}/\SiO{} for center wavelengths below about \qty{5}{\um}, and \aSi{} in combination with other metal oxides that exhibit lower loss at longer wavelengths, such as \TaO{}, for longer wavelengths. These perspectives highlight the scalability of the hybrid approach and warrant further process development for an even wider range of MIR applications.

\begin{backmatter}
\bmsection{Funding} Content in the funding section will be generated entirely from details submitted to Prism.

\bmsection{Acknowledgment} We thank our coauthors in Ref.~\cite{truong_mid-infrared_2023-1} for discussions and insights throughout the collaboration.

\bmsection{Disclosures} LWP: Thorlabs, Inc. (P), VW: Thorlabs, Inc. (P), UTOM AG (I,E), GWT: Thorlabs, Inc. (E), SCL: Thorlabs, Inc. (E), GDC: Thorlabs, Inc. (E,P), TS: UTOM AG (I).

\bmsection{Data availability} Data underlying the results presented in this paper are not publicly available at this time but may be obtained from the authors upon reasonable request.

\end{backmatter}

\bibliography{references_final}

\begin{thebibliography}{10}
\newcommand{\enquote}[1]{``#1''}

\bibitem{bai_optimizing_2011}
Y.~Bai, S.~R. Darvish, N.~Bandyopadhyay, \emph{et~al.}, \enquote{Optimizing facet coating of quantum cascade lasers for low power consumption,} {\protect\JournalTitle{J. Appl. Phys.}} \textbf{109}, 053103 (2011).

\bibitem{wysocki_widely_2008}
G.~Wysocki, R.~Lewicki, R.~Curl, \emph{et~al.}, \enquote{Widely tunable mode-hop free external cavity quantum cascade lasers for high resolution spectroscopy and chemical sensing,} {\protect\JournalTitle{Appl. Phys. B}} \textbf{92}, 305--311 (2008).

\bibitem{pecile_record-high_2024}
V.~F. Pecile, M.~Leskowschek, N.~Modsching, \emph{et~al.}, \enquote{Record-high power, low phase noise synchronously-pumped optical parametric oscillator tunable from 2.7 to 4.7 {{{\textmu}m}},} {\protect\JournalTitle{Appl. Phys. Lett.}} \textbf{125}, 231108 (2024).

\bibitem{adler_phase-stabilized_2009}
F.~Adler, K.~C. Cossel, M.~J. Thorpe, \emph{et~al.}, \enquote{Phase-stabilized, 1.5 {{W}} frequency comb at 2.8--4.8 {{{\textmu}m}},} {\protect\JournalTitle{Opt. Lett.}} \textbf{34}, 1330 (2009).

\bibitem{fleisher_optical_2017}
A.~J. Fleisher, D.~A. Long, Q.~Liu, \emph{et~al.}, \enquote{Optical measurement of radiocarbon below unity fraction modern by linear absorption spectroscopy,} {\protect\JournalTitle{J. Phys. Chem. Lett.}} \textbf{8}, 4550--4556 (2017).

\bibitem{terabayashi_optical_2017}
R.~Terabayashi, V.~Sonnenschein, H.~Tomita, \emph{et~al.}, \enquote{Optical feedback in {{DFB}} quantum cascade laser for mid-infrared cavity ring-down spectroscopy,} {\protect\JournalTitle{Hyperfine Interact.}} \textbf{238}, 10 (2017).

\bibitem{mccartt_room-temperature_2022}
A.~D. McCartt and J.~Jiang, \enquote{Room-temperature optical detection of {\textsuperscript{14}}{{CO}}{\textsubscript{2}} below the natural abundance with two-color cavity ring-down spectroscopy,} {\protect\JournalTitle{ACS Sens.}} \textbf{7}, 3258--3264 (2022).

\bibitem{kaariainen_optical_2019}
T.~K{\"a}{\"a}ri{\"a}inen and G.~Genoud, \enquote{Optical interruption of a quantum cascade laser for cavity ring-down spectroscopy,} {\protect\JournalTitle{Opt. Lett.}} \textbf{44}, 5294 (2019).

\bibitem{truong_mid-infrared_2023-1}
G.-W. Truong, L.~W. Perner, D.~M. Bailey, \emph{et~al.}, \enquote{Mid-infrared supermirrors with finesse exceeding 400 000,} {\protect\JournalTitle{Nat. Commun.}} \textbf{14}, 7846 (2023).

\bibitem{winkler_mid-infrared_2021-1}
G.~Winkler, L.~W. Perner, G.-W. Truong, \emph{et~al.}, \enquote{Mid-infrared interference coatings with excess optical loss below 10 {{ppm}},} {\protect\JournalTitle{Optica}} \textbf{8}, 686 (2021).

\bibitem{romanini_introduction_2014}
D.~Romanini, I.~Ventrillard, G.~M{\'e}jean, \emph{et~al.}, \enquote{Introduction to cavity enhanced absorption spectroscopy,} in \emph{Cavity-{{Enhanced Spectroscopy}} and {{Sensing}},}  G.~Gagliardi, H.-P. Loock, and {Rhodes, William T.}, eds. (Springer, 2014), pp. 1--60.

\bibitem{zhao_doppler-free_2020}
G.~Zhao, D.~M. Bailey, A.~J. Fleisher, \emph{et~al.}, \enquote{Doppler-free two-photon cavity ring-down spectroscopy of a nitrous oxide ({{N}}{$_{2}$}{{O}}) vibrational overtone transition,} {\protect\JournalTitle{Phys. Rev. A}} \textbf{101}, 062509 (2020).

\bibitem{delli_santi_biogenic_2021}
M.~G. Delli~Santi, S.~Bartalini, P.~Cancio, \emph{et~al.}, \enquote{Biogenic fraction determination in fuel blends by laser-based {\textsuperscript{14}}{{CO}}{\textsubscript{2}} detection,} {\protect\JournalTitle{Adv. Photonics Res.}} \textbf{2}, 2000069 (2021).

\bibitem{giusfredi_saturated-absorption_2010}
G.~Giusfredi, S.~Bartalini, S.~Borri, \emph{et~al.}, \enquote{Saturated-absorption cavity ring-down spectroscopy,} {\protect\JournalTitle{Phys. Rev. Lett.}} \textbf{104}, 110801 (2010).

\bibitem{giusfredi_theory_2015}
G.~Giusfredi, I.~Galli, D.~Mazzotti, \emph{et~al.}, \enquote{Theory of saturated-absorption cavity ring-down: Radiocarbon dioxide detection, a case study,} {\protect\JournalTitle{J. Opt. Soc. Am. B}} \textbf{32}, 2223 (2015).

\bibitem{zhao_frequency_2021}
G.~Zhao, J.~Tian, J.~T. Hodges, and A.~J. Fleisher, \enquote{Frequency stabilization of a quantum cascade laser by weak resonant feedback from a {{Fabry}}--{{Perot}} cavity,} {\protect\JournalTitle{Opt. Lett.}} \textbf{46}, 3057 (2021).

\bibitem{shumakova_short_2024}
V.~Shumakova and O.~H. Heckl, \enquote{A short guide to recent developments in laser-based gas phase spectroscopy, applications, and tools,} {\protect\JournalTitle{APL Photonics}} \textbf{9}, 010803 (2024).

\bibitem{jin_micro-fabricated_2022}
N.~Jin, C.~A. McLemore, D.~Mason, \emph{et~al.}, \enquote{Micro-fabricated mirrors with finesse exceeding one million,} {\protect\JournalTitle{Optica}} \textbf{9}, 965 (2022).

\bibitem{franta_optical_2016}
D.~Franta, D.~Ne{\v c}as, I.~Ohl{\'i}dal, and A.~Giglia, \enquote{Optical characterization of {{SiO}}{\textsubscript{2}} thin films using universal dispersion model over wide spectral range,} {\protect\JournalTitle{Proc. SPIE}} \textbf{9890}, 989014 (2016).

\bibitem{franta_wide_2025}
D.~Franta, J.~Voh{\'a}nka, J.~Dvo{\v r}{\'a}k, \emph{et~al.}, \enquote{Wide spectral range optical characterization of tantalum pentoxide ({{Ta}}{\textsubscript{2}}{{O}}{\textsubscript{5}}) films by the universal dispersion model,} {\protect\JournalTitle{Opt. Mater. Express}} \textbf{15}, 903 (2025).

\bibitem{chen_comparison_2020}
Y.~Chen, D.~Hahner, M.~Trubetskov, \emph{et~al.}, \enquote{Comparison of magnetron sputtering and ion beam sputtering on dispersive mirrors,} {\protect\JournalTitle{Appl. Phys. B}} \textbf{126}, 82 (2020).

\bibitem{habel_group_2016}
F.~Habel, M.~Trubetskov, and V.~Pervak, \enquote{Group delay dispersion measurements in the mid-infrared spectral range of 2-20 {{{\textmu}m}},} {\protect\JournalTitle{Opt. Express}} \textbf{24}, 16705 (2016).

\bibitem{cole_high-performance_2016}
G.~D. Cole, W.~Zhang, B.~J. Bjork, \emph{et~al.}, \enquote{High-performance near- and mid-infrared crystalline coatings,} {\protect\JournalTitle{Optica}} \textbf{3}, 647 (2016).

\bibitem{heiss_epitaxial_2001}
W.~Heiss, T.~Schwarzl, J.~Roither, \emph{et~al.}, \enquote{Epitaxial {{Bragg}} mirrors for the mid-infrared and their applications,} {\protect\JournalTitle{Prog. Quantum Electron.}} \textbf{25}, 193--228 (2001).

\bibitem{cole_laser-induced_2018}
G.~D. Cole, D.~Follman, P.~Heu, \emph{et~al.}, \enquote{Laser-induced damage measurements of crystalline coatings,} {\protect\JournalTitle{Proc. SPIE}} \textbf{10805}, 108051D (2018).

\bibitem{perner_simultaneous_2023}
L.~W. Perner, G.-W. Truong, D.~Follman, \emph{et~al.}, \enquote{Simultaneous measurement of mid-infrared refractive indices in thin-film heterostructures: Methodology and results for {{GaAs}}/{{AlGaAs}},} {\protect\JournalTitle{Phys. Rev. Research}} \textbf{5}, 033048 (2023).

\bibitem{cole_substrate-transferred_2025}
G.~D. Cole, V.~Wittwer, L.~W. Perner, \emph{et~al.}, \enquote{Substrate-transferred stacked optical coatings,} U.S. patent 12,352,987B2 (8 July 2025).

\bibitem{perner_ion-beam-sputtered_2025}
L.~W. Perner, V.~J. Wittwer, G.-W. Truong, \emph{et~al.}, \enquote{Ion-beam-sputtered mid-infrared coatings for hybrid supermirrors,} in \emph{Optical Interference Coatings (OIC),}  (2025). {{paper}} MC.4.

\bibitem{tatian_fitting_1984}
B.~Tatian, \enquote{Fitting refractive-index data with the {{Sellmeier}} dispersion formula,} {\protect\JournalTitle{Appl. Opt.}} \textbf{23}, 4477 (1984).

\bibitem{ciddor_refractive_1996}
P.~E. Ciddor, \enquote{Refractive index of air: New equations for the visible and near infrared,} {\protect\JournalTitle{Appl. Opt.}} \textbf{35}, 1566 (1996).

\bibitem{macleod_thin-film_2018}
H.~A. Macleod, \emph{Thin-Film Optical Filters\textup{, 5th ed.}} (CRC Press, 2018).

\bibitem{hirschfeld_focal_1978}
T.~Hirschfeld, \enquote{Focal shift photometric errors in fourier transform infrared spectroscopy,} {\protect\JournalTitle{Appl. Spectrosc.}} \textbf{32}, 508--509 (1978).

\bibitem{truong_near-infrared_2019}
G.~W. Truong, G.~Winkler, T.~Zederbauer, \emph{et~al.}, \enquote{Near-infrared scanning cavity ringdown for optical loss characterization of supermirrors,} {\protect\JournalTitle{Opt. Express}} \textbf{27}, 19141 (2019).

\end{thebibliography}

\end{document}